\documentclass[10pt,twocolumn]{revtex4}
\usepackage{graphicx,subfigure,epsfig,appendix}
\usepackage{booktabs,multirow,mathrsfs,amssymb,amsmath,amsthm}
\usepackage{algorithm,algorithmicx,framed,listings} 
\usepackage[colorlinks=true,linkcolor=blue,citecolor=blue,urlcolor=blue]{hyperref}
\usepackage[noend]{algpseudocode}
\newcommand{\bra}[1]{\langle #1|}
\newcommand{\ket}[1]{|#1\rangle}

\graphicspath{{figures/}}
\begin{document}

\title{Trainable Quantum Channels as Computational Primitives for Quantum Learning}

\author{Jingwei Wen$^{1}$}

\author{Runqing Zhang$^{1}$}
\author{Xusheng Xu$^{2}$}
\author{Xiaoxiao Hu$^{2}$}
\author{Jize Han$^{1}$}
\author{Ling Qian$^{1}$}
\author{Tiejun Wang$^{3}$}
\email{wangtiejun@bupt.edu.cn}
\author{Shijie Wei$^{4}$}
\email{weisj@baqis.ac.cn}
\author{Guilu Long$^{4,5,6,7}$}
\email{gllong@tsinghua.edu.cn}

\affiliation{$^{1}$ China Mobile (Suzhou) Software Technology Company Limited, Suzhou 215163, China}
\affiliation{$^{2}$ QuSpect Technology Company Limited, Beijing 100084, China}
\affiliation{$^{3}$ School of Physical Science and Technology, Beijing University of Posts and Telecommunications, Beijing 100876, China}
\affiliation{$^{4}$ Beijing Academy of Quantum Information Sciences, Beijing 100193, China}
\affiliation{$^{5}$ State Key Laboratory of Low-Dimensional Quantum Physics and Department of Physics, Tsinghua University, Beijing 100084, China}
\affiliation{$^{6}$ Frontier Science Center for Quantum Information, Beijing 100084, China}
\affiliation{$^{7}$ Beijing National Research Center for Information Science and Technology, Beijing 100084, China}

\begin{abstract}

Variational quantum learning is traditionally constrained to unitary dynamics, often treating quantum channels as detrimental noise. In this work, we reformulate the quantum channels as trainable computational primitives and establish a non-unitary quantum machine learning framework grounded in open-system dynamics. We demonstrate that the outputs of channel-enhanced quantum models form a structured superposition of multiple functional components. Each component is governed by an effective observable whose spectrum can be adaptively modulated during training, a significant departure from the spectral invariance in unitary transformations. Moreover, the proposed framework generalizes conventional unitary quantum models by retaining them as a special case while introducing additional non-unitary degrees of freedom. Furthermore, we reveal that trainable quantum channels enrich the optimization geometry through ensemble-averaged gradient and additional optimization directions induced by the Kraus operators. Extensive experiments on classification tasks using trainable amplitude-damping and phase-damping channels confirm enhanced optimization dynamics and predictive performance. In addition, we experimentally validate the proposed framework through hardware inference using ten-qubit quantum models implemented on the superconducting quantum processor, confirming its practical feasibility and hardware compatibility. Our work provides a principled approach for leveraging quantum channels as trainable resources and advances the design of high-performance quantum learning architectures.

\end{abstract}

\maketitle

\section{Introduction}

Quantum machine learning (QML) has emerged as one of the most important research directions in current quantum computing \cite{schuld2015introduction,biamonte2017quantum,schuld2019quantum}. Among various QML architectures, variational quantum neural networks (QNNs) stand out for their compatibility with noisy intermediate-scale quantum (NISQ) devices \cite{preskill2018quantum,bharti2022noisy}. Conventionally, given a data-encoded quantum state $\rho_x$, the QNNs process information through parameterized unitary operators $U(\theta)$, yielding the predictions as $f(x,\theta)=\mathrm{Tr}[OU(\theta)\rho_xU^\dagger(\theta)]$ with an observable $O$ and trainable parameter $\theta$ \cite{beer2020training,cerezo2021variational,alchieri2021introduction,wen2024enhancing}. Despite their strong theoretical expressivity, practical deployment of QNNs faces prominent challenges, including barren plateaus that cause vanishing gradients, persistent local minima that trap optimization, and performance degradation induced by hardware noise \cite{mcclean2018barren,kwak2021quantum,banchi2021generalization,cerezo2021cost,anschuetz2022quantum,wang2021noise,singkanipa2025beyond}.

Traditionally, the dissipation and quantum noise have been regarded as detrimental factors to the quantum learning \cite{wang2021noise,singkanipa2025beyond,cai2023quantum}. Extensive efforts have focused on the adverse effects of noise on model performance such as noise-induced barren plateaus and compromised generalization \cite{escudero2023assessing,kashif2024investigating,ahmed2025noisy,kashif2024nrqnn,njiki2026robustness}. For instance, systematic evaluations confirm that typical quantum noise channels, especially depolarizing noise, drastically reduce classification accuracy on the benchmark datasets like Iris \cite{kashif2024investigating}. To mitigate these issues, researchers have developed many strategies including noise-aware training \cite{wang2022quantumnat}, pre-training schemes \cite{scala2025improving} and optimized measurement observable selection \cite{kashif2025hqnet}, to help improve the performance of quantum learning processes. However, a paradigm shift is underway recently, re-evaluating noise not merely as a nuisance but as a potential computational resource \cite{domingo2023taking,zapusek2025dissipation}. A growing number of studies confirm that engineered dissipation can positively influence optimization landscapes and generalization \cite{sannia2024engineered,garcia2024effects,van2025exploiting,du2021quantum,huang2023certified,winderl2025constructing,ricci2026quantum,crognaletti2026estimates,rodriguez2026pac}. For example, it is found that the inclusion of properly designed Markovian losses after unitary circuits allows for the mitigation of barren plateaus \cite{sannia2024engineered}. It has also been proven that quantum noise can reshape the quantum Fisher information matrix, and activate previously inaccessible parameter directions, converting overparametrized QNNs into underparametrized models \cite{garcia2024effects}. Research reveals that bias-preserving noisy variational circuits can enhance the overall performance of variational quantum algorithms \cite{van2025exploiting}. Additionally, structured quantum noise has been verified to improve the adversarial robustness and privacy of QML models \cite{du2021quantum,huang2023certified,winderl2025constructing}. Specifically, Du et al. demonstrated that depolarizing noise could endow quantum classifiers with quantum differential privacy and strengthen their resistance to the adversarial attacks \cite{du2021quantum}. Huang et al. theoretically proved that injecting random noise into quantum rotation gates can improve overall model robustness \cite{huang2023certified}, while Winderl et al. constructed a universal quantum channel and validate their defense capability across multiple datasets \cite{winderl2025constructing}. 

Despite the advances, existing research treats quantum dissipation as an uncontrollable and passive environmental factor. A comprehensive investigation into actively adopting quantum channels as integrated and trainable modules for QML, encompassing model construction, theoretical performance analysis and experimental validation, has yet to be conducted.

To address these limitations, we propose a generalized quantum learning framework, dubbed channelQNNs, which incorporates circuit-based trainable quantum channels into variational QNNs. Instead of treating the loss of energy and information as static and unavoidable disturbances, we utilize parameterized completely positive trace-preserving (CPTP) maps as trainable non-unitary transformations within the learning architectures. We theoretically show that trainable quantum channels can enhance the quantum learning processes from both representational and optimization perspectives. In terms of representation, the model output forms a structured superposition of channel-induced functional components, each governed by an effective observable whose spectral properties can be adaptively modulated, in contrast to the spectral invariance imposed by unitary transformations. For optimization, the trainable quantum channels can enrich the optimization geometry through ensemble-averaged gradient dynamics over channel-evolved states and additional non-unitary optimization directions induced by channel parameterization. Notably, the proposed framework constitutes a generalization of conventional unitary quantum models and includes unitary models as a limiting case. Moreover, we implement the framework with trainable amplitude-damping (AD) and phase-damping (PD) channels, and conduct extensive evaluations on image classification and power grid stability prediction across different experimental setups. In addition, we experimentally demonstrate the proposed framework by deploying ten-qubit channelQNNs on the superconducting quantum processor for hardware inference, demonstrating its practical feasibility and compatibility with current quantum devices. Our results demonstrate that the trainable quantum channels can be harnessed as efficient computational resources, providing a promising route toward expressive and hardware-compatible quantum learning architectures.

\section{Results}

\begin{figure}
\centering
\includegraphics[width=\linewidth]{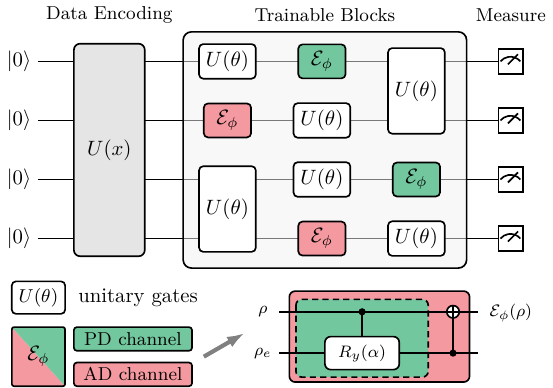}
\caption{\textbf{Trainable quantum channels in QNNs}. (a) The parameterized quantum channels $\mathcal{E}_{\phi}$ are incorporated into the variational architecture, with the channel parameters $\phi$ optimized jointly with the unitary parameters $\theta$ during the training process. The quantum-circuit implementation for the PD and AD channels is realized using an environmental qubit initialized as $\rho_e$ and controlled operations. The channel parameter is encoded through the controlled rotation $R_y(\alpha)$ according to the relation $\phi=\sin^2(\alpha/2)$.}
\label{fig1-channelQML}
\end{figure}

We here give the quantum circuit construction methods of QNNs architecture incorporating trainable quantum channels. As illustrated in the figure \ref{fig1-channelQML}, the proposed framework integrates parameterized non-unitary transformations into variational quantum learning, leading to the overall quantum model
\begin{equation}
f(\theta,\phi)=\mathrm{Tr}\left[O\mathcal{U}_{\theta}(\mathcal{E}_{\phi}(\rho_x))\right]
\label{eq:model}
\end{equation}
where the $\mathcal{U}_{\theta}(\rho)=U(\theta)\rho U^\dagger(\theta)$ denotes the variational unitary transformation, the $\mathcal{E}_{\phi}(\rho)$ is a CPTP channel parameterized by parameter $\phi\in[0,1]$, the $O$ is a Hermitian observable and $\rho_x$ is the data-related input quantum state. The trainable quantum channel can be represented with Kraus representation as
\begin{equation}
\mathcal{E}_{\phi}(\rho)=\sum_kE_k(\phi)\rho E_k^\dagger(\phi)
\end{equation}
where the parameter-dependent Kraus operators satisfy completeness condition $\sum_kE_k^\dagger(\phi)E_k(\phi)=I$ with $I$ denoting the identity operator \cite{tong2006kraus,nielsen2010quantum}. 

Unlike conventional open-system quantum learning, the channel parameter $\phi$ is treated as a variational parameter here and optimized with the unitary parameters $\theta$ during training process. We focus on two representative quantum channels, namely AD and PD channels \cite{xin2017quantum,wei2018efficient}, which characterize energy relaxation and dephasing processes, respectively. Their dynamics can be implemented through parameterized unitary operators with one ancillary environment qubit as
\begin{equation}
\mathcal{E}_{\phi}(\rho)=\mathrm{Tr}_{e}[V(\phi)(\rho\otimes\rho_{e})V^\dagger(\phi)]
\end{equation}
where the environment qubit is initialized in the quantum state $\rho_{e}=\ket{0}\bra{0}$. Specifically, the corresponding operators for the PD and AD channels are given by $V_{pd}(\phi)=\ket{0}\bra{0}\otimes I+\ket{1}\bra{1}\otimes R_y(2\arcsin\sqrt{\phi})$ and $V_{ad}(\phi)=[I\otimes\ket{0}\bra{0}+\sigma_x\otimes\ket{1}\bra{1}]V_{pd}(\phi)$, where $\sigma_x$ is the Pauli matrix \cite{nielsen2010quantum,mickelson2026quantum}. This ancilla-assisted construction realizes the non-unitary channel evolution through unitary dynamics in an enlarged Hilbert space while avoiding post-selection procedures. More importantly, it enables trainable quantum channels to be seamlessly incorporated into variational quantum circuits and optimized jointly with unitary transformations. We detail the quantum-circuit construction of trainable AD and PD channels in the Methods section.

\section{Theoretical Analysis}

In this part, we provide a theoretical characterization of trainable quantum channels within variational quantum learning. We show that the proposed framework improves quantum learning from two perspectives. First, the trainable quantum channels extend the hypothesis space of conventional unitary quantum models by inducing a structured superposition of channel-dependent predictors. This extension generalizes the unitary QNNs while preserving them as a special case. Second, the trainable quantum channels can fundamentally modify the optimization geometry through ensemble average of channel-evolved states, and introduce additional non-unitary gradient pathways.

\subsection{Channel-Induced Expansion of Quantum Hypothesis Spaces}

\begin{figure*}
\centering
\includegraphics[width=\linewidth]{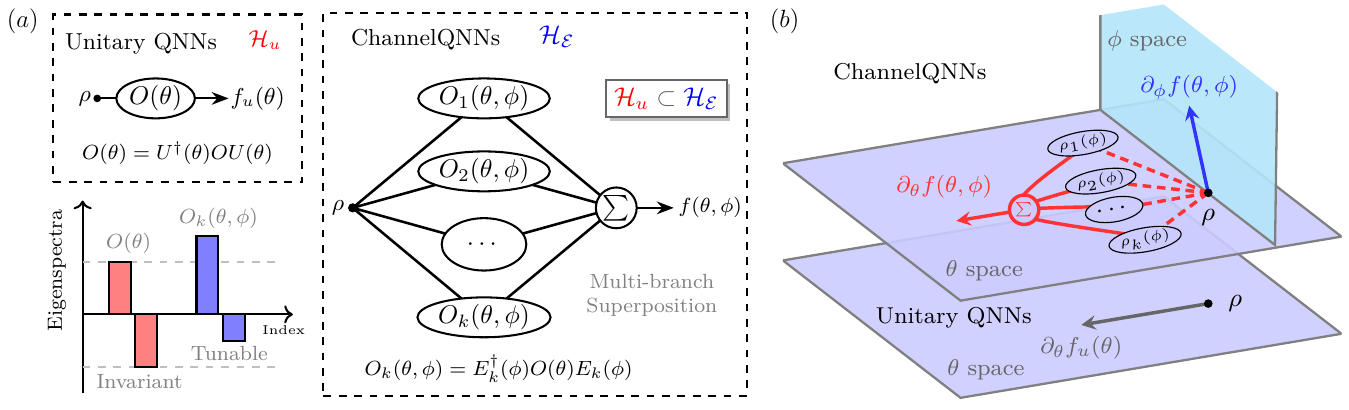}
\caption{\textbf{Functional spaces and optimization geometry of unitary QNNs and channelQNNs}. (a) Functional space comparison. The hypothesis spaces of unitary QNNs and channelQNNs are denoted by $\mathcal{H}_u$ and $\mathcal{H}_\mathcal{E}$, respectively. The unitary QNNs span $\mathcal{H}_u$ with spectrum-invariant observables $O(\theta)=U^\dagger(\theta)OU(\theta)$, while channelQNNs support multi-branch superposition of tunable observables $O_k(\theta,\phi)=E_k^\dagger(\phi)O(\theta)E_k(\phi)$ over a larger hypothesis space $\mathcal{H}_\mathcal{E}$ with $\mathcal{H}_u \subset \mathcal{H}_\mathcal{E}$. (b) Geometric visualization of gradients for optimization. The unitary QNNs access a single gradient $\partial_\theta f_u(\theta)$ confined to the $\theta$ parameter space. For channelQNNs, multiple sub-gradients are generated from distinct Kraus-branch evolved states $\rho_k(\phi)=E_k(\phi)\rho E_k^\dagger(\phi)$ and ensemble-averaged into the gradient $\partial_\theta f(\theta,\phi)$, together with an additional gradient $\partial_\phi f(\theta,\phi)$ along the channel parameter direction.}
\label{fig2-channelQML}
\end{figure*}

We consider the hypothesis space generated by variational quantum circuits equipped with trainable quantum channels as
\begin{equation}
\mathcal{H}_{\mathcal{E}}=\left\{f(\theta,\phi) =\mathrm{Tr}\left[O\mathcal{U}_{\theta}(\mathcal{E}_{\phi}(\rho))\right] | \{\theta,\phi\} \in \mathbb{R}\right\}
\end{equation}
where the $\mathcal{E}_{\phi}$ denotes a parameterized CPTP map with trainable channel parameters $\phi$. Using the Kraus representation and linearity of the trace, the quantum model output can be uniformly written as
\begin{equation}
f(\theta,\phi)=\sum_{k}\mathrm{Tr}\big(E_k^\dagger(\phi)O(\theta)E_k(\phi)\rho\big)
\end{equation}
where the parameterized observable is defined as $O(\theta)=U^\dagger(\theta)OU(\theta)$. We further define the channel-deformed observables as
\begin{equation}
O_k(\theta,\phi)=E_k^\dagger(\phi)O(\theta)E_k(\phi)
\end{equation}
where the effective observable $O_k(\theta,\phi)$ is still Hermitian, but the eigenvalues can be modulated by the parameters $\phi$ and $\theta$. Consequently, we can obtain a unified representation for the quantum model output as
\begin{equation}
f(\theta,\phi)=\sum_{k}\mathrm{Tr}\big(O_k(\theta,\phi)\rho\big)=\sum_{k}f_k(\theta,\phi)
\label{loss-channel}
\end{equation}
where $f_k(\theta,\phi)=\mathrm{Tr}(O_k(\theta,\phi)\rho)$ denotes the contribution associated with the $k$-th Kraus branch. For comparison, the output of a conventional unitary quantum model is
\begin{equation}
f_u(\theta)=\mathrm{Tr}\big(OU(\theta)\rho U^\dagger(\theta)\big)=\mathrm{Tr}\big(O(\theta)\rho\big)
\label{loss-unitary}
\end{equation}
which corresponds to a single observable trajectory generated by unitary evolution. 

It can be concluded that there exists a fundamental distinction from conventional unitary quantum models: \textit{Instead of being generated by a single unitary trajectory, the output of the trainable-channel quantum model can be expressed as a structured superposition of multiple channel-dependent predictors}. Each branch corresponds to a distinct observable deformation induced by the associated Kraus operator. Such a branching structure enriches the hypothesis space by generating multiple functional components, while the Kraus completeness condition imposes correlations among different branches.

Moreover, there is another critical distinction between the observable $O(\theta)$ of unitary models and channel-induced observables $O_k(\theta,\phi)$. The unitary operations constitute a similarity transformation and cannot change the spectrum of observables. In contrast, when incorporating the trainable quantum channels via Kraus operators $E_k(\phi)$, the eigenvalues of the channel-deformed observable $O_k(\theta,\phi)$ are generally not related to $O$ through similarity transformations. Their spectral structure becomes dependent on the trainable channel parameters $\phi$. As a result, \textit{the trainable quantum channels introduce adaptive observable geometries that cannot be generated by unitary dynamics}. This additional degree of freedom enables the model to realize richer functional representations and provides a mechanism for extending the approximation capability. 

Furthermore, an important property of the proposed framework is that it naturally contains unitary QNNs. The channel actions of parameterized quantum channels $\mathcal{E}_{\phi}(\rho)$ would continuously approach the identity map when the channel parameter $\phi$ vanishes, which means $\lim_{\phi\to0}\mathcal{E}_{\phi}(\rho) = \rho$. It indicates that the hypothesis spaces of unitary quantum models and the proposed quantum models satisfy 
\begin{equation}
\mathcal{H}_{u}=\{f_u(\theta)=\lim_{\phi\to0}f(\theta,\phi)| \theta \in \mathbb{R}\}\subset \mathcal{H}_{\mathcal{E}}
\end{equation}
which confirms that \textit{the trainable-channel quantum models constitute a generalization of conventional unitary architectures}. It implies that if the optimal solution of a learning task resides within the unitary manifold, the trainable channel parameters $\phi$ can converge to zero during training to recover the unitary dynamics. Therefore, our framework does not sacrifice the well-established capabilities of unitary QNNs, but rather extends them by offering a broader hypothesis space where the non-unitary transformations are indispensable. We summarize the above mechanisms of enhanced expressivity for channelQNNs in the figure \ref{fig2-channelQML}\textcolor{blue}{(a)}, and provide detailed derivations of the characteristics of the hypothesis spaces under trainable AD and PD channels in the Methods section.

\subsection{Enriched Optimization Geometry through Trainable Channels}

Beyond representation, the trainable quantum channels can also modify the optimization geometry of variational quantum learning. For conventional unitary models, consider a parameterized gate $U(\theta)=\exp(-iH\theta)$ with a generator Hamiltonian $H$, the gradient of output in equation (\ref{loss-unitary}) relative to parameter $\theta$ is 
\begin{equation}
\partial_{\theta} f_u(\theta)=-i\mathrm{Tr}([O(\theta), H ]\rho)
\end{equation}
The optimization dynamics are therefore governed by the commutator-induced unitary transformations. For the trainable-channel models in equation (\ref{loss-channel}), two distinct gradient mechanisms emerge. On the one hand, the gradient in the unitary direction is
\begin{equation}
\partial_{\theta} f(\theta,\phi)=-i\sum_{k}\mathrm{Tr}\big([O(\theta), H]E_k(\phi)\rho E_k^\dagger(\phi)\big)\label{gradient-theta}
\end{equation}
where we define the branch quantum states as $\rho_k(\phi)=E_k(\phi)\rho E_k^\dagger(\phi)$ corresponding to the action of the $k$-th Kraus operator. It can be concluded that while the gradient with respect to the unitary parameters, $\partial_{\theta} f(\theta,\phi)$, preserves the same commutator-based form as unitary QNNs, \textit{it is effectively averaged over an ensemble of channel-evolved branches rather than being evaluated on a single quantum state.} So the trainable quantum channels can fundamentally enrich the optimization geometry compared to conventional unitary models.

On the other hand, the trainable channels introduce additional optimization directions associated with the channel parameters as
\begin{equation}
\partial_{\phi} f(\theta,\phi)=\sum_{k}\mathrm{Tr}(A_k(\theta,\phi)\rho)
\label{gradient-phi}
\end{equation}
where the derivative of effective observable is $A_k(\theta,\phi) = \partial_{\phi} O_k(\theta,\phi)=\partial_{\phi}E_k^\dagger(\phi)O(\theta)E_k(\phi)+E_k^\dagger(\phi)O(\theta)\partial_{\phi}E_k(\phi)$. Unlike the commutator structure in unitary optimization, \textit{the trainable channel offers a distinct mechanism for optimization, where the gradient is governed by the derivative of the Kraus operators}. Consequently, the optimization landscape is no longer restricted to a purely unitary parameter manifold. Instead, optimization proceeds simultaneously along unitary directions $\theta$ and non-unitary channel directions $\phi$. This enlarged optimization geometry provides additional pathways for navigating complex loss landscapes in a fundamentally different mechanism: $\partial_\theta f(\theta,\phi) = 0$ requires $\mathrm{Tr}\big([O(\theta),H]\,\rho_k(\phi)\big) = 0$ for all $k$, whereas $\partial_\phi f(\theta,\phi) = 0$ requires $\mathrm{Tr}\big(A_k(\theta,\phi)\,\rho\big) = 0$ for all $k$. These two conditions are governed by distinct matrix elements of the density matrix, and their zero sets generically do not intersect. By transforming the optimization problem relying on a single gradient mechanism into a scheme with dual gradient mechanisms, channelQNNs can mitigate the risk of stagnation within unfavorable optimization landscapes. We summarize the above mechanisms of enriched optimization geometry for channelQNNs in the figure \ref{fig2-channelQML}\textcolor{blue}{(b)}, and present the gradient characteristics under the trainable AD and PD channels in the Methods section.

\section{Experimental Evaluation}

\subsection{Numerical Validation}

To evaluate the effectiveness of quantum learning models with trainable quantum channels, we conduct binary classification experiments on two representative datasets: the MNIST handwritten digit dataset \cite{deng2012mnist} and the Electrical Grid Stability Simulated Dataset (EGSSD) \cite{arzamasov2018electrical}. Throughout the experiments, we compare three different quantum models including unitary QNNs, QNNs augmented with trainable PD channels (PD-QNNs), and QNNs augmented with trainable AD channels (AD-QNNs). 

\begin{figure}
    \centering
    \includegraphics[width=\linewidth]{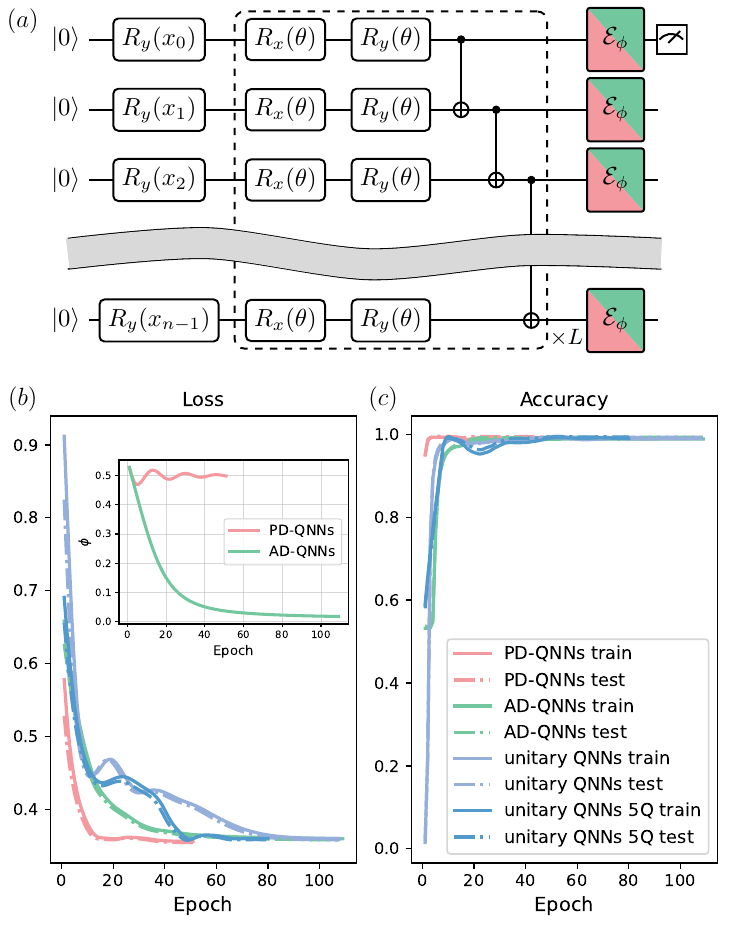}
    \caption{\textbf{Quantum circuit architecture and algorithm performance on the MNIST binary classification task} (a) The variational quantum circuit with feature encoding, $L$ repeated ansatz layers, and optional non-unitary channel $\mathcal{E}_\phi$. (b,c) The loss and accuracy on the training and test dataset, comparing the four-qubit unitary QNNs, PD-QNNs and AD-QNNs with one trainable channel on the first qubit, together with five-qubit unitary QNNs. The insert panel shows the evolution of channel parameter $\phi$.}
    \label{fig2-MNIST4qubit}
\end{figure}

\begin{figure}
    \centering
    \includegraphics[width=\linewidth]{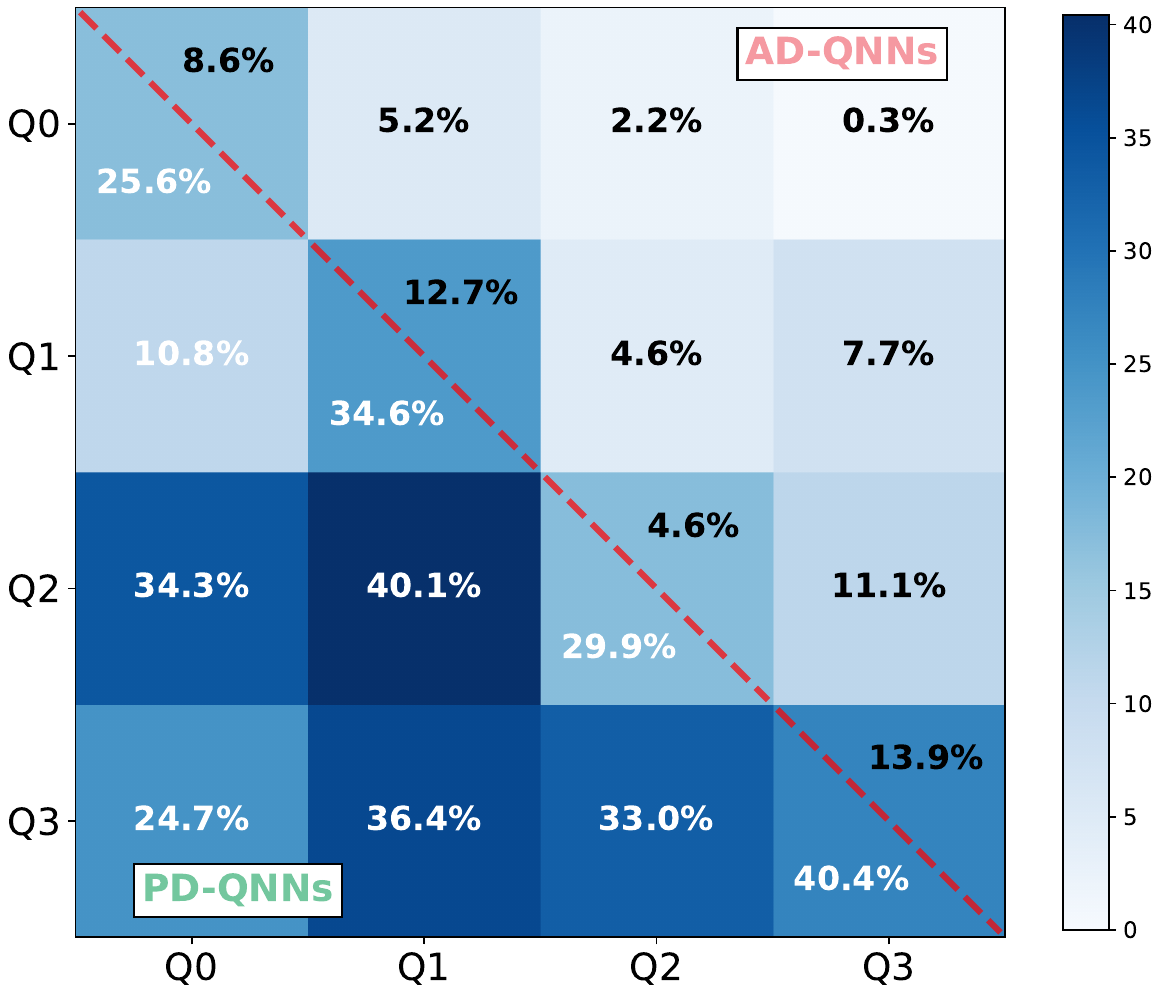}
    \caption{\textbf{Average ratio of saved training epochs relative to the unitary baseline for trainable quantum channels}. All values are averaged over three independent repeated runs. The data at off-diagonal position ($Q_i$,$Q_j$) represent the results for applying two AD (upper triangle) or two PD (lower triangle) channels on qubit $i$ and qubit $j$. Every diagonal cell is partitioned by a black diagonal separator: the top-right (bottom-left) sub-triangle shows the results of implementing one AD (PD) channel on individual qubit $i$.}
      \label{fig3-MNISTepoch}
\end{figure}

\begin{figure*}
    \centering
        \includegraphics[width=\linewidth]{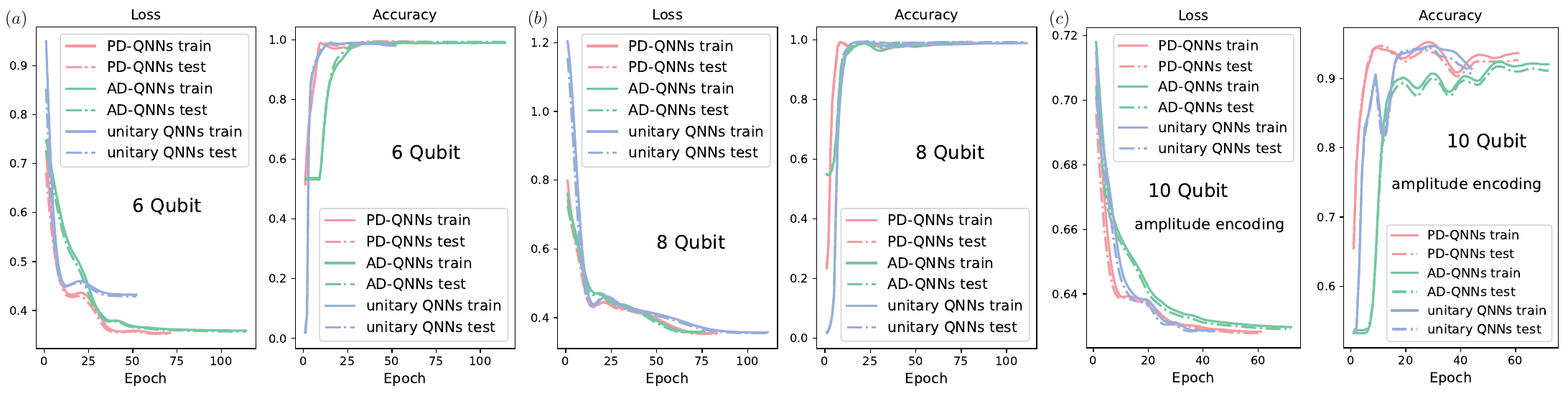}
    \caption{\textbf{Training and test performance comparison of unitary QNNs, PD-QNNs, and AD-QNNs on the MNIST binary classification task across varying qubit numbers and encoding schemes.} (a) $n=6$-qubit models with angle encoding; (b) $n=8$-qubit models with angle encoding; (c) $n=10$-qubit models with amplitude encoding. }
    \label{fig4-MNISTvary}
\end{figure*}

As shown in the figure \ref{fig2-MNIST4qubit}\textcolor{blue}{(a)}, the input data $x_i$ with label $y_i$ is embedded into an $n$-qubit quantum model via single-qubit rotation gates defined as $R_y(x_i) = \exp(-i\sigma_y x_i/2)$. The variational block $U({\theta})$ consists of $L$ stacked layers as $U(\theta) = \prod_{l=1}^{L} U_l({\theta}^{(l)})$. Each layer is composed of parameterized single-qubit rotations followed by nearest-neighbor two-qubit entangled gates, which is formulated as
\begin{equation}
U_{l}(\theta^{(l)}) =  \prod_{k=0}^{n-2} \textup{CNOT}(k,k+1)\bigotimes_{i=0}^{n-1} R_y(\theta_2^{(i)})R_x(\theta_1^{(i)})
\label{ansatz1}
\end{equation}
In total, the quantum circuit contains $2nL$ trainable rotation parameters and $(n-1)L$ nearest-neighbor CNOT gates. For the channel-enhanced architectures, the trainable quantum channel is inserted between the variational quantum blocks and measurement operators. At the end of circuit, we measure the expectation value of the Pauli-$Z$ operator on the first qubit, and map it to a classification probability via $\hat{y}_i = \sigma(\langle \sigma_z\rangle)$.

We train all models for 200 epochs using cross-entropy loss function with $L_2$ regularization.
The loss function is defined as
\begin{equation}
\mathcal{L} = -\frac{1}{N}\sum_{i=1}^N \left[ y_i \log \hat{y}_i + (1-y_i)\log(1-\hat{y}_i) \right] + \frac{\lambda}{|\theta|}\sum_j \theta_j^2
\end{equation}
where $N$ is the number of samples, $\lambda=10^{-3}$ is the regularization strength, and $|\theta|$ represents the total number of trainable parameters. Quantum models are optimized using the Adam optimizer with a cosine-decay learning-rate schedule. To improve training efficiency, early stopping is activated when the variance of training losses over the most recent ten epochs falls below $10^{-8}$. All the experiments are implemented using PennyLane \cite{bergholm2018pennylane}.

We first consider a four-qubit experiment with $L=2$ layers and one channel operator on the first qubit using MNIST dataset, where handwritten digits 0 and 1 were selected as the classification task, and the training set contains 12665 samples and the test set contains 2115 samples. For the angle-encoding experiments, principal component analysis (PCA) \cite{abdi2010principal} is applied to reduce the original $28\times28$ images into a four-qubit quantum state $\ket{x}=\bigotimes_{i=1}^{4}R_y(x_i)\ket{0}$ \cite{rath2024quantum}. It can be concluded from figure \ref{fig2-MNIST4qubit}\textcolor{blue}{(b,c)} that all models achieve nearly perfect classification accuracy, since the PCA-reduced binary task is relatively simple. Nevertheless, substantial differences emerge in the optimization processes. The two channel-augmented QNNs exhibit smoother convergence than unitary baselines, benefiting from richer gradient trajectories. The PD-QNNs converges far faster and its damping parameter stabilizes at a non-zero value, indicating the model actively leverages phase damping process. By contrast, the damping strength of AD-QNNs gradually decays toward zero, requiring nearly identical epoch to the unitary baseline. This distinct evolution demonstrates the adaptive characteristic of our framework: the model automatically adjusts dissipation intensity according to task demands. We further quantitatively quantify the acceleration brought by trainable quantum channels by calculating the proportion of training epochs saved relative to the unitary model, as illustrated in the figure \ref{fig3-MNISTepoch}. All reported values are averaged over three independent repeated training. We can conclude that trainable channels applied to single-qubit and two-qubit via both AD and PD schemes reduce the total number of training epochs needed for convergence. Across all tested experimental setups, PD channels deliver an average epoch reduction of 31.0\%, whereas AD channels yield an average acceleration of 7.1\%. This result indicates that trainable quantum channels can serve as powerful computational primitives to reshape the optimization landscape, which substantially accelerate optimization and smooth the loss landscape while preserving classification accuracy. Moreover, we offer a five-qubit unitary model to exclude qubit number as a contributing factor to performance promotion. Instead, the performance superiority verifies the beneficial role of dynamically adjustable dissipation embedded within the variational architecture. 

Furthermore, we increase the system size to six and eight qubits under angle encoding with one trainable channel on the first qubit. The results in the figure \ref{fig4-MNISTvary}\textcolor{blue}{(a,b)} demonstrate that the optimization advantage of trainable quantum channels persists as the number of qubits increases. In both settings, PD-QNNs and AD-QNNs achieve lower final loss, or exhibit faster and smoother optimization trajectories than the unitary baseline. Moreover, we evaluate the proposed framework under amplitude encoding \cite{rath2024quantum}. The input data are normalized and embedded directly into a $10$-qubit quantum state with $L=3$ layers, omitting PCA and other dimensionality-reduction preprocessing. The corresponding results are presented in the figure \ref{fig4-MNISTvary}\textcolor{blue}{(c)}, and we can find that channel models have higher and more stable accuracy throughout training, while maintaining excellent convergence stability. Collectively, the MNIST experiments indicate that the proposed trainable-channel architectures consistently outperform conventional unitary QNNs across different qubit numbers and encoding strategies.

\begin{figure}
    \centering
    \includegraphics[width=\linewidth]{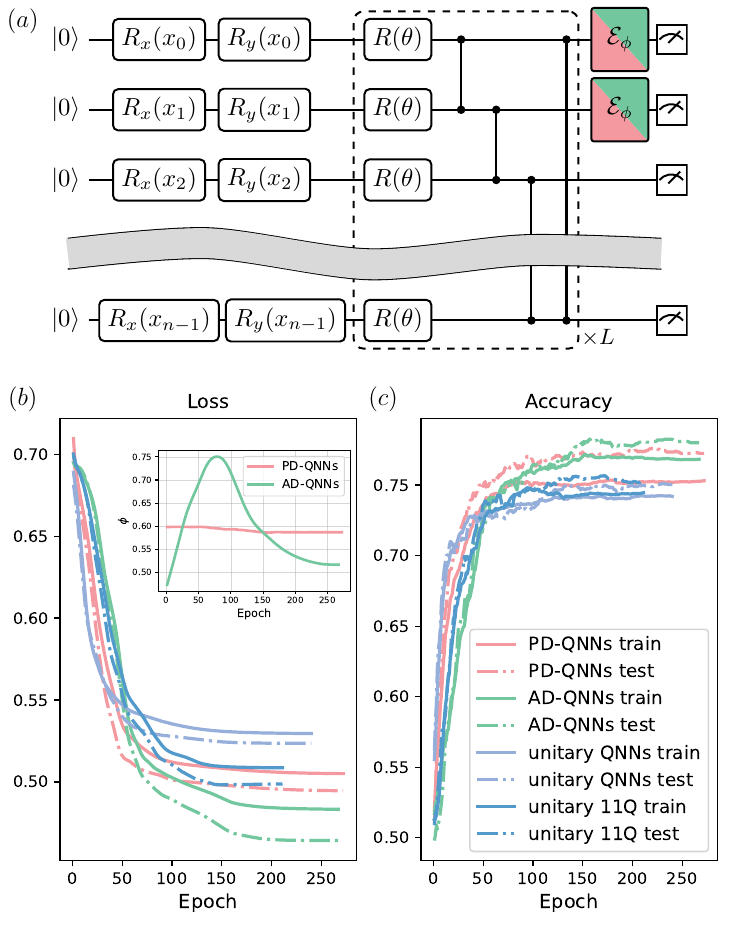}
     \caption{\textbf{Quantum circuit architecture and algorithm performance on the EGSSD binary classification task} (a) The variational quantum circuit with feature encoding, $L$ repeated ansatz layers, and optional non-unitary channel $\mathcal{E}_\phi$. (b,c) The loss and accuracy on the training and test dataset, comparing the ten-qubit unitary QNNs, PD-QNNs and AD-QNNs with one trainable channel on the first qubit, together with eleven-qubit unitary QNNs. The insert panel shows the evolution of channel parameter $\phi$.}
    \label{fig5-EGSSD}
\end{figure}

\begin{figure}
    \centering
    \includegraphics[width=\linewidth]{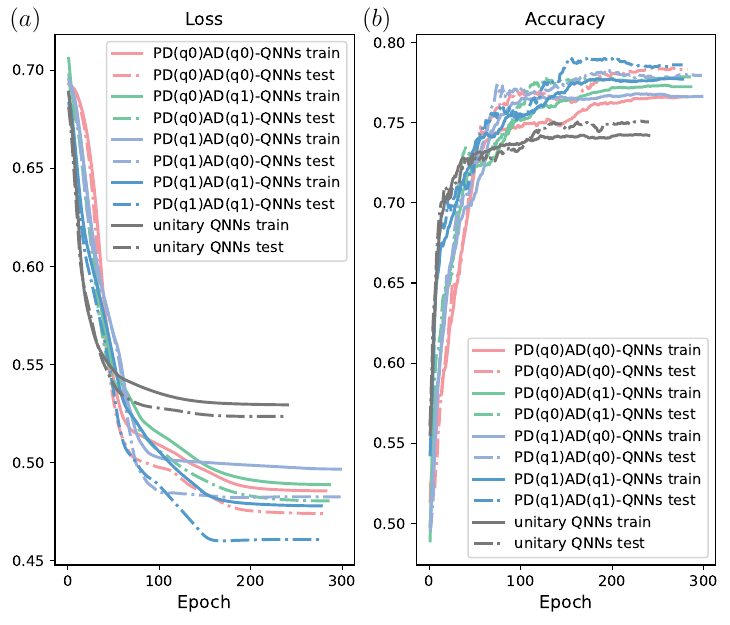}
     \caption{\textbf{Convergence curves of loss and classification accuracy on the EGSSD with hybrid AD and PD channels}. The loss (a) and accuracy (b) for the ten-qubit unitary baseline QNNs and four hybrid QNNs equipped with combined AD and PD non-unitary channels on different qubit positions. The notation PD($q_i$)AD($q_j$)-QNNs denotes the quantum model equipped with PD channel on qubit $i$ and AD channel on qubit $j$.}
    \label{fig6-EGSSDepoch}
\end{figure}

To further evaluate the proposed framework on a more challenging real-world task, we consider a classification task on EGSSD. This dataset contains eleven-dimensional features describing the operating conditions of a four-node electrical grid and is labeled according to whether the grid remains stable or not. Following standard practice, 8000 samples are used for training and 2000 samples are reserved for testing. After PCA preprocessing, the data are encoded into a ten-qubit quantum circuit using angle encoding, as shown in the figure \ref{fig5-EGSSD}\textcolor{blue}{(a)}. The variational quantum classifier contains three trainable layers composed of universal single-qubit rotational gates $R(\theta)=R_z(\theta_{1})R_y(\theta_2)R_z(\theta_3)$ and ring-shaped controlled-$Z$ entangling operations. The model output is obtained from the global observable $O=\bigotimes_{i=1}^{10} Z_i$, followed by a sigmoid activation function for binary classification. The experimental results of loss and classification accuracies are shown in the figure \ref{fig5-EGSSD}\textcolor{blue}{(b,c)}. Compared with MNIST, this task exhibits a considerably more complicated decision boundary, making the differences between architectures more apparent. The unitary model converges to the highest final loss and achieves the lowest accuracy. In contrast, both PD-QNNs and AD-QNNs converge to significantly lower loss values and achieve consistently higher accuracy throughout training, indicating superior performance. The evolution of the trainable channel parameter $\phi$ reveals the distinct optimization behaviors induced by different channels. For PD-QNNs, the value of $\phi$ slowly declines and finally stabilizes, establishing moderate dissipation regularization. In contrast, AD-QNNs demonstrate a more active regulation process: its $\phi$ first rises and then gradually decays toward a stationary value. Such dynamics enable the adaptive mechanism to flexibly modulate energy dissipation to match the intricate decision boundary of task, which explains the more favorable optimization performance of AD-QNNs. Furthermore, to isolate the effect of trainable channels from the influence of qubit number, we additionally compare against an eleven-qubit unitary QNN. Relative to the ten-qubit unitary baseline, expanding the qubit numbers of unitary baseline can merely improve test accuracy by 0.2\%, whereas the ten-qubit models embedded with PD and AD channels obtain accuracy improvement of 2.2\% and 3.0\%. Such contrast confirms that our structured non-unitary designs offer a far more effective path to performance gains than merely increasing qubit resources.

Moreover, we evaluate the performance of quantum circuits equipped with hybrid AD and PD channels, as shown in the figure \ref{fig6-EGSSDepoch}. It can be found that all configurations with dual AD-PD channels imposed on different qubit pairs yield consistent positive gains, which converge to lower loss and higher prediction accuracy than the unitary baseline. Across different qubit allocation schemes, the dual-channel model PD($q_1$)AD($q_1$)-QNNs with both PD and AD channels applied to qubit 1 achieves the largest accuracy gain of $3.9\%$ over the unitary baseline, while the other three hybrid configurations yield moderate improvements: $2.8\%$ for PD($q_0$)AD($q_1$)-QNNs, $3.2\%$ for PD($q_1$)AD($q_0$)-QNNs, and $3.3\%$ for PD($q_0$)AD($q_0$)-QNNs. The results demonstrate that combining AD and PD trainable channels on designated qubits can yield consistent performance gains.

\subsection{Hardware Demonstration}

\begin{figure}
    \centering
    \includegraphics[width=\linewidth]{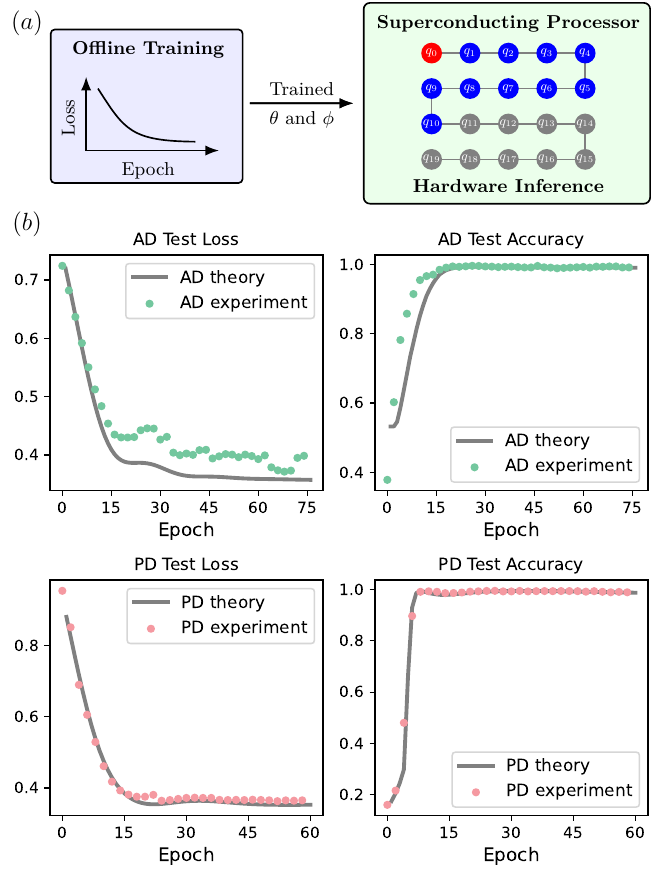}
  \caption{\textbf{Experimental results of loss and accuracy for the PD- and AD-enhanced QNNs models on quantum hardwares}. (a) Schematic workflow and topological structure of quantum chips, where ten qubits ($q_1$-$q_{10}$) are used as the quantum learning models and one ancillary qubit ($q_{0}$) as the environment system. (b) Comparison between theoretical simulation and hardware experimental results. Solid gray lines denote theoretical predictions, while the dots represent experimental data.}
      \label{fig7-MNISTexp}
\end{figure}

To further validate the practical feasibility of the proposed algorithms, we implement the channel-enhanced QNNs on a superconducting quantum processor, and the parameter details of quantum chips are provided in the Methods section. Following the above experiment setups, the MNIST dataset after PCA is encoded by angle encoding and both the AD and PD channels are implemented using quantum architectures in the figure \ref{fig2-MNIST4qubit}\textcolor{blue}{(a)} with $n=10$ qubits and $L=1$ layer. Rather than performing variational optimization directly on hardware, we adopt an inference-oriented workflow to reduce quantum resources while preserving the complete physical realization of trainable channels. Specifically, the variational parameters are first optimized through classical simulation, and then the optimized parameters are deployed onto the superconducting processor to reproduce the inference process. The quantum processors will be called 1024 times to obtain measurement values. To evaluate the consistency between numerical predictions and hardware execution, we reconstruct the inference performance using the optimized parameters throughout the training process, thereby experimentally validating the complete optimization trajectory. As shown in the figure \ref{fig7-MNISTexp}, the hardware implementation achieves classification accuracies close to numerical simulations. To quantitatively characterize the performance degradation between ideal simulation and hardware measurements, we employ the mean absolute error (MAE) as the evaluation metric. Given $N$ epochs, let $M_{s,i}$ and $M_{e,i}$ denote the target metric (accuracy or loss) from simulation and hardware inference at epoch $i$. The MAE is defined as $\Delta M =\sum_{i=1}^{N}\vert M_{s,i} - M_{e,i}\vert/N$, which captures the average magnitude of performance mismatch. Specifically, the MAE for loss values reach 0.038 and 0.017 for the AD and PD channel models, respectively, while the corresponding average accuracy degradations are 2.0\% (AD) and 1.3\% (PD). These small discrepancies are mainly attributed to hardware imperfections, including gate errors, readout errors, and decoherence, while the overall inference performance remains highly consistent with the numerical expectations. These results indicate that the proposed trainable quantum channels can be efficiently realized on current superconducting quantum processors and provide a hardware-compatible route toward practical channel-enhanced quantum learning.

\section{Conclusion}

In summary, we have proposed a generalized variational quantum learning framework that elevates quantum channels to trainable computational primitives grounded in open-system dynamics. By incorporating the parameterized quantum channels into optimization processes of QNNs, we demonstrate theoretically that the proposed architecture enables structured superpositions of functional components governed by adaptively modulated observables, transcending the spectral rigidity inherent to purely unitary transformations. The channel-enhanced framework can serve as a generalization of conventional unitary models, which arise as a natural limiting case. Furthermore, it enriches the optimization geometry through ensemble-averaged gradient dynamics and supplementary gradient pathways introduced by quantum channels. Taken together, the trainable quantum channels can transform the quantum learning model from a single-observable learner into a structured ensemble of channel-conditioned predictors, while introducing adaptive observable spectra and additional optimization directions. Numerical experiments on quantum classification tasks using trainable AD and PD channels validate the effectiveness of our approach, exhibiting improved optimization dynamics and predictive performance. Moreover, we experimentally validate the proposed framework by deploying ten-qubit quantum learning models on the superconducting quantum processor for hardware inference. The hardware results closely reproduce the numerical predictions, demonstrating the practical feasibility and hardware compatibility of trainable quantum channels. Our work establishes a principled paradigm for harnessing dissipation and quantum channels as valuable computational resources, paving the way for the development of high-performance and hardware-compatible quantum architectures.

We note that Yao et al. recently proposed a variational dissipative quantum framework dedicated to ground-state search and state recovery \cite{yao2026variational}. Distinct from their research scope, our work focuses on general QML tasks and establishes a systematic theoretical analysis and experimental validations for the channel-based quantum learning models.

\section{Methods}

\subsection{Quantum-Circuit Realization of Trainable Quantum Channels}

The trainable quantum channels considered in this work are implemented through an explicit interaction between the system qubits and ancillary environment qubits. Such a construction follows the Stinespring dilation theorem, which states that any CPTP map can be realized as a unitary evolution on a larger Hilbert space followed by tracing out the environment \cite{mickelson2026quantum}.

Specifically, we introduce an ancillary qubit initialized in the quantum state $\rho_e=\ket{0}\bra{0}$
and perform a parameterized unitary interaction $V(\phi)$ between the system and environment, where the trainable channel parameter $\phi$ is encoded through the system-environment interaction. 

For the AD channel as shown in figure \ref{fig1-channelQML}, the unitary $V_{ad}(\phi)$ is composed of a controlled rotation $R_y(\alpha)=\exp(-i\sigma_y\alpha/2)$ followed by a CNOT gate. Consider an arbitrary system qubit with state $|\psi\rangle=a|0\rangle+b|1\rangle$, the quantum state after evolution is
\begin{equation}
|\Psi_{ad}\rangle=\big(a|0\rangle+b\cos\frac{\alpha}{2}|1\rangle\big)\ket{0}_{e}+b\sin\frac{\alpha}{2}|0\rangle\ket{1}_{e}
\end{equation}
Then the final quantum state of work system can be obtained by tracing out the environment as
\begin{equation}
\begin{split}
\rho_{ad}^{f}&
=\sum_{i,j=0,1}\big(\bra{i} \otimes I\big)\ket{\Psi_{ad}}\bra{\Psi_{ad}}\big(I\otimes \ket{i}\big)\\
&=\begin{bmatrix}
|a|^2+|b|^2\sin^2(\alpha/2) &a b^*\cos(\alpha/2)\\
a^* b\cos(\alpha/2)&|b|^2\cos^2(\alpha/2)
\end{bmatrix}
\label{VthetaAD}
\end{split}
\end{equation}

Correspondingly, the Kraus operators for the AD channel are 
\begin{equation}
E_0^{ad}(\phi) = \begin{bmatrix} 1 & 0 \\ 0 & \sqrt{1-\phi} \end{bmatrix}, \quad
E_1^{ad}(\phi) = \begin{bmatrix} 0 & \sqrt{\phi} \\ 0 & 0 \end{bmatrix}
\end{equation}
where $\phi\in[0,1]$ denotes the decay probability of the quantum channel. Then the output state is
\begin{equation}
\begin{split}
\mathcal{E}_{\phi}^{ad}(\ket{\psi}\bra{\psi})&=\sum_{k=0,1}E_k^{ad}(\phi)\ket{\psi}\bra{\psi} (E_k^{ad}(\phi))^\dagger\\
=&\begin{bmatrix}
|a|^2+\phi|b|^2&ab^*\sqrt{1-\phi}\\
a^*b\sqrt{1-\phi}&(1-\phi)|b|^2
\end{bmatrix}
\end{split}
\label{KrausAD}
\end{equation}
Therefore, the quantum circuit can realize an AD process by encoding the damping strength as $\phi=\sin^2(\alpha/2)$, unifying equations \eqref{VthetaAD} and \eqref{KrausAD}.

For the PD channel, the CNOT operation is removed and the state is $|\Psi_{pd}\rangle=\big(a|0\rangle+b\cos(\alpha/2)|1\rangle\big)\ket{0}_{e}+b\sin(\alpha/2)|1\rangle\ket{1}_{e}$ after quantum circuit. Then the final quantum state of work system becomes
\begin{equation}
\begin{split}
\rho_{pd}^{f}&
=\sum_{i,j=0,1}\big(\bra{i} \otimes I\big)\ket{\Psi_{pd}}\bra{\Psi_{pd}}\big(I\otimes \ket{i}\big)\\
&=\begin{bmatrix}
|a|^2&a b^*\cos(\alpha/2)\\
a^* b\cos(\alpha/2)&|b|^2
\end{bmatrix}
\label{VthetaPD}
\end{split}
\end{equation}
And the Kraus operators for the PD channel are
\begin{equation}
E_0^{pd}(\phi) = \begin{bmatrix} 1 & 0 \\ 0 & \sqrt{1-\phi} \end{bmatrix}, \quad
E_1^{pd}(\phi) = \begin{bmatrix} 0 & 0 \\ 0 & \sqrt{\phi} \end{bmatrix}
\end{equation}
which results the output quantum state as
\begin{equation}
\begin{split}
\mathcal{E}_{\phi}^{pd}(\ket{\psi}\bra{\psi})&=\sum_{k=0,1}E_k^{pd}(\phi)\ket{\psi}\bra{\psi} (E_k^{pd}(\phi))^\dagger\\
=&\begin{bmatrix}
|a|^2&ab^*\sqrt{1-\phi}\\
a^*b\sqrt{1-\phi}&|b|^2
\end{bmatrix}
\end{split}
\label{KrausPD}
\end{equation}
Therefore, from equations \eqref{VthetaPD} and \eqref{KrausPD}, the same relation between the quantum-circuit parameter and damping strength as derived for the AD channel holds for the PD channel, namely $\phi=\sin^2(\alpha/2)$, which enables the circuit-based implementation of PD channels.

The above derivation reveals that the proposed trainable quantum channels are not heuristic noise injections but exact physical realizations of CPTP maps obtained from system-environment interactions. The ancillary qubit serves as an explicit environmental degree of freedom, while the trainable parameter $\phi$ controls the strength of energy and information exchange between the system and environment. Consequently, the proposed framework transforms environmental dissipation from an uncontrollable source of noise into a trainable computational primitive that can be optimized jointly with the variational circuit parameters. Notably, the operations are implemented via unitary evolution on the enlarged system-environment Hilbert space without requiring post-selection or ancillary measurement.

\subsection{Hypothesis spaces under AD and PD channels}

\begin{figure}
\centering
\includegraphics[width=\linewidth]{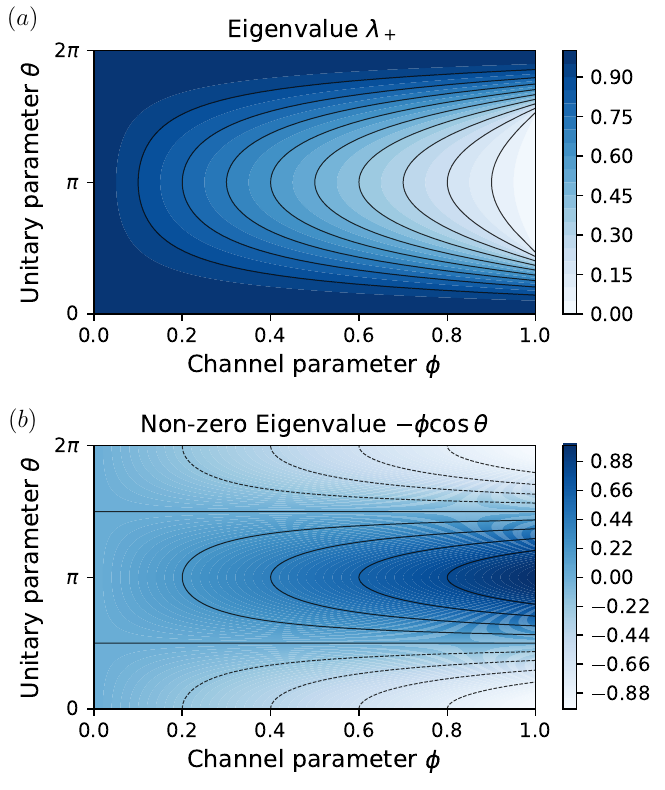}
\caption{\textbf{Eigenvalues of the effective observable as functions of channel and unitary parameters.} The panel (a) shows the positive eigenvalue $\lambda_+$, while the panel (b) shows the non-zero eigenvalue $-\phi\cos\theta$.}
\label{fig7-eigen}
\end{figure}

To concretely demonstrate the mechanism of hypothesis space expansion and spectral modulation, we analyze the AD and PD channels with a single-qubit observable $O(\theta) = \boldsymbol{n}(\theta) \cdot \boldsymbol{\sigma}$, where the $\boldsymbol{n}(\theta) = (\sin\theta, 0, \cos\theta)$ is a unit vector on the Bloch sphere and $\boldsymbol{\sigma} = (\sigma_x, \sigma_y, \sigma_z)$ denotes the vector of Pauli operators. The parameterized quantum channels are represented by Kraus operators $E_k(\phi)$.

The AD channel models the energy dissipation process, and the effective observables $O_k^{ad}(\theta,\phi)= (E_k^{ad}(\phi))^\dagger O(\theta) E_k^{ad}(\phi)$ for $k=0$ is
\begin{equation}
\begin{split}
O_0^{ad}(\theta,\phi)= \begin{bmatrix} \cos\theta & \sin\theta\sqrt{1-\phi} \\ \sin\theta\sqrt{1-\phi} & -(1-\phi)\cos\theta \end{bmatrix}
\end{split}
\end{equation}
and the eigenvalues are
\begin{equation}
\lambda_\pm^{(0)}(\theta,\phi) = \frac{\phi\cos\theta\pm \sqrt{\phi^2\cos^2\theta + 4(1-\phi)}}{2}
\end{equation}
As $\phi$ increases from $0$ to $1$, the eigenvalues shrink from $\pm 1$ to $\{0,\cos\theta\}$. The spectrum is compressed, unlike the fixed spectrum of unitary operations. Similarly, the effective observables for $k=1$ is $O_1^{ad}(\theta,\phi)= \phi \cos\theta(I-\sigma_z)/2$ with eigenvalues $\{0,\phi \cos\theta\}$, where the non-zero eigenvalue scales linearly with the damping strength $\phi$.

Moreover, the PD channel models loss of phase coherence, and the effective observable $O_0^{pd}(\theta,\phi)=O_0^{ad}(\theta,\phi)$ for $k=0$ case, and $O_1^{pd}(\theta,\phi)= \phi \cos\theta(\sigma_z-I)/2$ with eigenvalues $\{0,-\phi \cos\theta\}$ for $k=1$ case. The eigenvalues of the effective observable act as a function of the channel parameter $\phi$ and the optimization parameter $\theta$, as shown in the figure \ref{fig7-eigen}.

Furthermore, the output for the AD and PD channels is the sum of these branches $f^{ad/pd}(\theta,\phi) =\sum_{k=0,1} \mathrm{Tr}[O_k^{ad/pd}(\theta,\phi)\rho]$, which reveals that the model implements a structured superposition of sub-functions. Crucially, since the eigenvalues of $O_k(\theta,\phi)$ are continuous functions of the channel parameter $\phi$ and unitary optimization parameter $\theta$, the model possesses dynamically tunable spectra, allowing it to adapt its representational geometry beyond the constraints of unitary manifolds. 

\subsection{Gradients under AD and PD channels}

Given the single-qubit effective observable $O(\theta) = \sin\theta \sigma_x + \cos\theta \sigma_z$ and the Hamiltonian $H =\sigma_z/2$, the commutator is $[O(\theta), H] = -i\sin\theta \sigma_y$, resulting the unitary gradient as
\begin{equation}
\partial_{\theta} f_u(\theta) = -\sin\theta \cdot \mathrm{Tr}(\sigma_y \rho) = -2\sin\theta \,\mathrm{Im}(\rho_{01})
\end{equation}
where the elements of density matrix are $\rho_{ij}=\bra{i}\rho\ket{j}$ with relations $\rho_{ij}=\rho_{ji}^{*}$ and $\rho_{00}+\rho_{11}=1$. The gradient for the parameter $\theta$ under quantum channels is averaged over different channel branches as
\begin{equation}
\begin{split}
\partial_{\theta} f(\theta,\phi) &= -\sin\theta \sum_{k=0,1} \mathrm{Tr}\big(\sigma_y E_k(\phi)\rho E_k^\dagger(\phi)\big) \\
&= -\sin\theta\big(\langle\sigma_y\rangle_{\rho_0(\phi)}+\langle\sigma_y\rangle_{\rho_1(\phi)}\big)
\end{split}
\end{equation}

For the both the AD and PD channels, we have $\langle\sigma_y\rangle_{\rho_0(\phi)}=2\sqrt{1-\phi}\,\mathrm{Im}(\rho_{01})$ and $\langle\sigma_y\rangle_{\rho_1(\phi)}=0$, then the gradient is
\begin{equation}
\begin{split}
\partial_{\theta} f^{ad/pd}(\theta,\phi)=-2\sin\theta\sqrt{1-\phi}\,\mathrm{Im}(\rho_{01})
\end{split}
\end{equation}

We then turn to the gradient in quantum channels for parameter $\phi$. For the AD channel, the Kraus derivatives yield $A_0^{ad}(\theta,\phi) =-\sin\theta\sigma_x/(2\sqrt{1-\phi})+\cos\theta(I-\sigma_z)/2$ and $A_1^{ad}(\theta,\phi) =\cos\theta(I-\sigma_z)/2$. Thus, the gradient is
\begin{equation}
\partial_{\phi} f^{ad}(\theta,\phi)= -\frac{\sin\theta}{\sqrt{1-\phi}}\mathrm{Re}(\rho_{01}) + 2\cos\theta\rho_{11}
\end{equation}
 For the PD channel, we can have $A_0^{pd}(\theta,\phi) = A_0^{ad}(\theta,\phi)$ and $A_1^{pd}(\theta,\phi) =-\cos\theta(I-\sigma_z)/2$, then the resulting gradient is
\begin{equation}
\partial_{\phi} f^{pd}(\theta,\phi)= -\frac{\sin\theta}{\sqrt{1-\phi}}\mathrm{Re}(\rho_{01})
\end{equation}

These calculations confirm that the gradients on unitary parameter $\partial_{\theta} f_u(\theta)$ and $\partial_\theta f(\theta,\phi)$ are driven by $\mathrm{Im}(\rho_{01})$, while the channel gradient $\partial_\phi f(\theta,\phi)$ is driven by $\mathrm{Re}(\rho_{01})$ and $\rho_{11}$ via the non-commutative deformation of the observable. The two gradients depend on distinct matrix elements of the density matrix and generally form structurally independent optimization directions on the parameter manifold.

\subsection{Quantum Hardware}

The experiments are conducted on the Z02 superconducting quantum processor, a 20-qubit quantum chip fabricated by Hangzhou Logical Qubit Technology Company Limited \cite{logicalqubit}. Each qubit is capacitively coupled to its nearest-neighbors, forming a one-dimensional chain structure. We utilize eleven physical qubits ($q_0$--$q_{10}$) to implement our channelQNNs algorithm. Specifically, ten qubits $q_1$--$q_{10}$ act as the working system, while qubit $q_0$ serves as an ancillary qubit to realize the AD and PD channels.

The characteristic parameters of the selected physical qubits are summarized in the Table~\ref{tab:z01_device_parameters}. Here, $T_1$ denotes the energy-relaxation time, $T_{2,\mathrm{E}}$ denotes the coherence time measured using a Hahn-echo sequence, and $T_2^{*}$ denotes the Ramsey dephasing time. $\mathcal{F}_{\mathrm{XEB}}$ represents the circuit fidelity estimated using cross-entropy benchmarking, whereas $F_{00}$ and $F_{11}$ denote the probabilities of correctly identifying the prepared $\left|0\right\rangle$ and $\left|1\right\rangle$ states, respectively. $F_{cz}^{j,j+1}$ represents the fidelity of CZ gate between the qubits $q_{j}$ and $q_{j+1}$. Across the eleven qubits used in this work, the average values of $T_1$, $T_{2,\mathrm{E}}$, and $T_2^{*}$ are $83.20\mu\mathrm{s}$, $7.21\mu\mathrm{s}$, and $1.31\mu\mathrm{s}$, respectively. The average XEB fidelity is $99.82\%$, while the average readout-assignment fidelities $F_{00}$ and $F_{11}$ are $96.60\%$ and $94.01\%$, respectively. 

\begin{table*}
    \centering
    \caption{Characteristic parameters of the eleven physical qubits $q_0$--$q_{10}$ used to implement the channelQNNs models on the Z01 superconducting quantum processor.}
    \label{tab:z01_device_parameters}
    \setlength{\tabcolsep}{2.5mm}{
     \begin{tabular}{cccccccccccc} %
        \hline
        \hline
        Parameter
        & $q_0$ & $q_1$ & $q_2$
        & $q_3$ & $q_4$ & $q_5$
        & $q_6$ & $q_7$ & $q_8$
        & $q_9$ & $q_{10}$ \\
        \hline
        $T_1$ $(\mu\mathrm{s})$
        & 72.95 & 120.71 & 82.87 & 46.18 & 76.31 & 81.93
        & 59.32 & 99.21 & 96.32 & 100.53 & 66.97 \\
        
        $T_{2,\mathrm{E}}$ $(\mu\mathrm{s})$
        & 7.57 & 9.38 & 6.87 & 5.57 & 7.49 & 9.05
        & 3.55 & 5.83 & 6.56 & 6.35 & 6.35 \\

        $T_2^{*}$ $(\mu\mathrm{s})$
        & 0.79 & 0.98 & 0.91 & 1.03 & 2.38 & 1.86
        & 1.08 & 1.09 & 1.53 & 1.30 & 1.44 \\
        
        $\mathcal{F}_{\mathrm{XEB}}$ $(\%)$
        & 99.90 & 99.92 & 99.84 & 99.57 & 99.84 & 99.92
        & 99.80 & 99.87 & 99.84 & 99.85 & 99.92 \\

        $F_{00}$
        & 0.977 & 0.952 & 0.975 & 0.974 & 0.974 & 0.938
        & 0.954 & 0.979 & 0.974 & 0.965 & 0.967 \\

        $F_{11}$
        & 0.966 & 0.978 & 0.969 & 0.916 & 0.943 & 0.932
        & 0.965 & 0.905 & 0.958 & 0.956 & 0.930 \\
        
        $F_{cz}^{j,j+1}$
        & 0.992 & 0.993 & 0.986 & 0.975 & 0.993 & 0.994 
        & 0.991 & 0.988 & 0.994 & 0.992 & - \\
        \hline
        \hline
    \end{tabular}%
    }
\end{table*}


\section{Acknowledgements}
We are grateful to Logicalqubit for furnishing hardware resources that facilitated this research.


\bibliographystyle{unsrtnat}
\bibliography{wen_channelQNNs.bib}


\end{document}